\documentclass[twocolumn,showpacs,preprintnumbers,amsmath,amssymb,prl]{revtex4-1}
\usepackage{amsmath,amssymb}
\usepackage{graphicx}
\usepackage{verbatim}

\usepackage[dvips,colorlinks=true,bookmarks=false,citecolor=blue,urlcolor=blue]{hyperref} %latex w/dvips

\usepackage{lmodern}
\usepackage[T1]{fontenc}
\usepackage[latin9]{inputenc}
\usepackage{color}
\usepackage{amssymb}
\usepackage{graphicx}
\usepackage{esint}
\usepackage{verbatim}

\begin{document}

\title{Extraordinary localization of TE-waves at the graphene layer}

\author{Yuliy V. Bludov$^{1}$}
\author{Daria A. Smirnova$^{2}$}
\author{Yuri S. Kivshar$^{2}$}
\author{N. M. R. Peres$^{1}$}
\author{Mikhail I. Vasilevskiy$^{1}$}

\affiliation{$^{1}$Centro de Física and Departamento de Física, Universidade
do Minho, Campus de Gualtar, Braga 4710-057, Portugal \\
$^{2}$Nonlinear Physics Center, Research School of Physics and Engineering,
Australian National University, Canberra ACT 0200, Australia}

\begin{abstract}
The propagation of electromagnetic waves along the surface of a nonlinear dielectric covered
by a graphene layer is investigated. The main result is that such a surface can support and stabilize
nonlinear transverse electric (TE) plasmon polaritons. We demonstrated that these nonlinear TE modes
have a subwavelength localization in the direction perpendicular to the surface, with the intensity much
higher than that of the incident wave which excites the polariton.
\end{abstract}


\maketitle

\textit{Introduction.}
Nonlinear plasmonics is still a young, but fast growing,  research field~\cite{natphot}.
It encompasses both the nonlinear response of the active medium --the metal-- and
that of the surrounding dielectric. Nonlinear response of plasmonic systems has been observed
both in metal films and in metallic nanostructures~\cite{natphot}. In the present context, from
many nonlinear effects allowed by nonlinear optics, second-harmonic generation and
self-action Kerr effect have been the most studied. Kerr effect refers to the modification of
the refractive index of a system by the electric field. In this case, the nonlinear
susceptibility depends on the intensity of the electric field.

Plasmonics in graphene~\cite{novnatphoton} is a recent and intense field of research, impelled by
many theoretical proposals~\cite{c:primer,theor1,theor3,theor4} and experimental results~\cite{Basovexp,Koppens_exp}.
In particular, it has been shown that graphene supports $p-$polarized surface plasmon polaritons, or transverse magnetic (TM) surface waves~\cite{theor1,theor3,theor4}, with subwavelength localization in the direction perpendicular to the surface.
Contrary to an ordinary metal, it has  been shown that graphene also supports TE-type electromagnetic
surface waves~\cite{Mikhailov2007}, in a well defined and narrow frequency window.
This novel type of surface waves exist in graphene as a consequence that the imaginary part of its inter-band
 optical conductivity may become negative. The existence of such type of waves was invoked to explain, e.g.
 the broadband polarizing effect of graphene~\cite{Bao2011}. The TE waves can also be important in
 multi-layer structures~\cite{our_JETPL}. Unfortunately, the degree of localization of the TE-type surface plasmon--polaritons around the graphene layer is very weak, with the electromagnetic field behaving almost as free radiation.

\begin{figure}[b]
\includegraphics[clip,width=8.5cm] {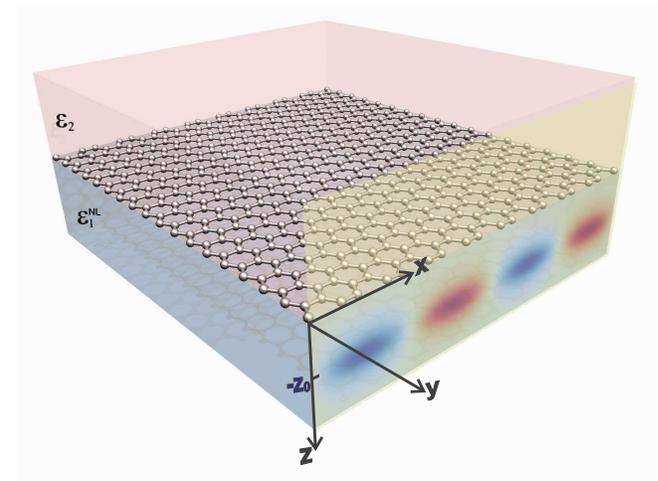}  % {fig1_graphen}  % {fig1_KB}
\caption{(Color online) Schematics of a graphene layer separating the upper linear
and lower nonlinear dielectric media with the dielectric permittivities $\varepsilon_2$
and $\varepsilon_1^{\text{NL}}$, respectively, shown with
a spatial profile of the nonlinear surface polariton.}
\label{fig:geometry}
\end{figure}

In the realm of linear optics there is no hope that the degree of localization of
the TE-type electromagnetic wave can be enhanced. Fortunately, nonlinear optics rescue us
from this limitation. The idea is relatively simple. For a TE-wave, the solutions of
Maxwell's equation in the Kerr regime support localized fields, described by hyperbolic
functions. Then, a simple experimental setup can be envisioned: a graphene sheet
is cladded by two dielectrics, one being linear (for example, air) and the
other nonlinear (for example, some special type of polymer). Electromagnetic radiation
is let to impinge from the linear dielectric onto graphene and a surface wave is excited (e.g. with the help of a
microfabricated grating). Due to the presence of the nonlinear dielectric
underneath, the field is localized in the vicinity of the interface. Furthermore, the enhancement of the
electric field associated with the formation of the surface wave further enhances
the nonlinear effect. Moreover, as we will see below, the presence of the nonlinear dielectric also frees us
from the narrow frequency window predicted by the linear theory for the existence of
TE-waves. Indeed, it is shown that this type of waves can exist even if the imaginary part of
the conductivity is positive. The main results of this Letter are: (i) subwavelength localization of TE-type
waves around the graphene-covered surface or interface;
(ii) enlarged frequency window for the existence of the TE-waves;
 (iii) strong enhancement of the localized field relatively to its value in the incident wave;
(iv) stability of the TE-type nonlinear waves, which have a soliton character.
% \begin{enumerate}
% \item sub-wavelength localization of TE-waves;
% \item enlarged energy window for the existence of TE-waves;
% \item strong enhancement of the localized field relatively to its value in free space;
% \item stability of the TE-waves, which have a soliton character.
% \end{enumerate}
All these aspects are promoted by the presence of the nonlinear dielectric in the vicinity of graphene.
In what follows
we will see how these results appear from a simple formulation of the nonlinear problem.
%------------------------------------------------------------------------------------------------------------------------------------------------

\textit{Model and Equations.} We consider a graphene sheet covering a flat interface
(at $z=0$) between two semi-infinite dielectric media (see Fig.~\ref{fig:geometry}).
The upper dielectric has a linear dielectric permittivity $\varepsilon_{2}>0$, and the {\em nonlinear} substrate has a
dielectric permittivity $\varepsilon_{1}^{\text{NL}}=\varepsilon_{1}+\chi^{(3)}|E|^{2}$; then we are exploiting here the Kerr effect.
The parameter $\chi^{(3)}>0$ is the nonlinear coefficient of the self-focusing medium, which
for some types of polymers can have values as large as $\chi^{(3)}=1.4\cdot10^{-18}\,\mathrm{m}^{2}/\mathrm{V}^{2}$.
We assume that the substrate and the capping dielectric occupy the half-spaces $z>0$ and $z<0$, respectively (see Fig.~\ref{fig:geometry}).
Moreover, we suppose that the incident electromagnetic radiation, in the form
of TE-waves, is uniform along the direction $y$, with the field vectors given by $\vec{E}=\left\{ 0,E_{y},0\right\}$
and $\vec{H}=\left\{ H_{x},0,H_{z}\right\}$. The temporal dependence of the electromagnetic field has the standard
form $\sim\exp(-i\omega t)$, where $\omega$ is the frequency.

Maxwell's equations for a TE-wave can be written in the form
\begin{eqnarray*}
&&\frac{\partial H_{x}^{(j)}}{\partial z}-\frac{\partial H_{z}^{(j)}}{\partial x}=-i\omega\varepsilon_{0}\left[\varepsilon_{j}+\delta_{j,1}\chi^{(3)}\left|E_{y}^{(j)}\right|^{2}\right]E_{y}^{(j)},  %\label{eq:max1-s}
\\
&&\frac{\partial E_{y}^{(j)}}{\partial z}=-i\omega\mu_{0}H_{x}^{(j)}, \quad % \label{eq:max2-s}
\frac{\partial E_{y}^{(j)}}{\partial x}=i\omega\mu_{0}H_{z}^{(j)}\,; \quad j=1,2\,. % \label{eq:max3-s}
\end{eqnarray*}
Introducing  the dimensionless and
slowly varying amplitudes of the electromagnetic field, ${\cal E}_{y}^{(j)}(x,z)=\left(\chi^{(3)}/2\right)^{1/2}E_{y}^{(j)}\exp(-ik_{x}x)$ and
${\cal H}_{x,z}^{(j)}(x,z)=\left(\chi^{(3)}/2\right)^{1/2}c\mu_{0}H_{x,z}^{(j)}\exp(-ik_{x}x)$,
it follows from the above equations that
\begin{eqnarray}
2ik_{x}\frac{\partial{\cal E}_{y}^{(j)}}{\partial x}+\frac{\partial^{2}{\cal E}_{y}^{(j)}}{\partial z^{2}}-k_{x}^{2}{\cal E}_{y}^{(j)}+\label{eq:wave-eq}\\
\left(\frac{\omega}{c}\right)^{2}\left[\varepsilon_{j}+2\delta_{j,1}\left|{\cal E}_{y}^{(j)}\right|^{2}\right]{\cal E}_{y}^{(j)}=0\:,\nonumber
\label{eq_Max_non_linear}
\end{eqnarray}
where $k_{x}$ is the in-plane component of the wavevector and $\delta_{j,m}$ is the Kronecker symbol.

\textit{Nonlinear Surface Modes.}
In the nonlinear medium ($j=1$), the stationary ($x$-independent) solution of wave equation (\ref{eq:wave-eq}) can be represented in the form of localized modes.
Indeed the solution has the simple form,
\begin{eqnarray}
{\cal {\cal E}}_{y}^{(1)}(z) & = & \frac{cp_{1}}{\omega}\frac{1}{\cosh\left[p_{1}\left(z+z_{0}\right)\right]},\label{eq:E1}\\
{\cal {\cal H}}_{x}^{(1)}(z) & = & \frac{c^{2}p_{1}^{2}}{i\omega^{2}}\frac{\sinh\left[p_{1}\left(z+z_{0}\right)\right]}{\cosh^{2}\left[p_{1}\left(z+z_{0}\right)\right]}\,.\label{eq:H1}
\end{eqnarray}
It is clear that the wave is localized around $z=-z_0$; this type of solution is sometimes called a spatial soliton.
The parameter $z_0$ can be either positive or negative. We are most interested in the case where
the TE-wave is localized in the nonlinear medium, for which $z_{0}<0$.

On the other hand, in the linear medium ($j=2$) the solution has the standard form
\begin{eqnarray}
{\cal {\cal E}}_{y}^{(2)}(z) & = & {\cal {\cal E}}_{y}(0)\exp\left(p_{2}z\right),\label{eq:E2}\\
{\cal {\cal H}}_{x}^{(2)}(z) & = & -\frac{cp_{2}}{i\omega}{\cal {\cal E}}_{y}(0)\exp\left(p_{2}z\right)\,.
\label{eq:H2}
\end{eqnarray}
In the above, we have introduced the parameter $p_{j}^{2}=k_{x}^{2}-\left(\omega/c\right)^{2}\varepsilon_{j}$.
The electric field at the interface, ${\cal {\cal E}}_{y}(0)$, and the dispersion relation for the
TE-waves are determined from the boundary conditions. These are
${\cal E}_{y}^{(2)}(0)={\cal E}_{y}^{(1)}\left(0\right)$
and ${\cal H}_{x}^{(1)}(0)-{\cal H}_{x}^{(2)}(0)=c\mu_{0}\sigma\left(\omega\right){\cal E}_{y}^{(2)}\left(0\right)$,
from which follows the dispersion relation:
\[
p_{1}\tanh(p_{1}z_{0})+p_{2}=i\omega\mu_{0}\sigma\left(\omega\right),
\]
or, equivalently,
\begin{equation}
s\left[p_{1}^{2}-\left(\frac{\omega}{c}\right)^{2}\left|{\cal E}_{y}\left(0\right)\right|^{2}\right]^{1/2} + p_{2}=i\omega\mu_{0}\sigma\left(\omega\right),
\label{eq:dr1}
\end{equation}
where $s=\pm1$ stands for the sign of the parameter $z_{0}$.
We note that the  dispersion relation of the nonlinear surface polaritons is determined
by the graphene conductivity $\sigma\left(\omega\right)$,
which has been given, for example, in Ref. \onlinecite{c:primer} and depends on the Fermi energy, $\mu$.
The general trend of the optical conductivity of graphene is as follows:
in the low-frequency range, the Drude term exceeds significantly
the inter-band contribution, for both the real and imaginary parts, while in the high-frequency range (that is, close to twice the Fermi energy) the latter contribution dominates. Moreover, in the vicinity of the frequency $\omega=2\mu$ the real part of the graphene conductivity increases drastically and achieves the universal conductivity value, $\sigma_{0}=\pi e^{2}/(2h)$ ($h$ is the Planck constant), whereas the imaginary
part, which is negative, attains its minimum value.
Also, there exists a special frequency $\omega_{*}$, where the imaginary part of the conductivity, $\sigma^{\prime\prime}=\mathrm{Im}\left[\sigma\left(\omega\right)\right]$,
vanishes, i.e. $\sigma^{\prime\prime}\left(\omega_{*}\right)=0$.

Without the graphene layer, the TE-wave could exist only in the case $s=-1$ and for a {\it single} value of the electric field amplitude at the interface~\cite{Tomlinson1980}, determined by $\left|{\cal E}_{y}\left(0\right)\right|^{2}=\varepsilon_{2}-\varepsilon_{1}\equiv-\Delta_{0}$ (for $\Delta_{0}<0$).
On the contrary, and most importantly, in the case where a graphene layer is present, the existence range
for the TE-wave becomes significantly wider. Indeed, Eq. (\ref{eq:dr1}) can be rewritten in terms of $\Delta=\varepsilon_{1}-\varepsilon_{2}+\left|{\cal E}_{y}\left(0\right)\right|^{2}\equiv \Delta_{0}+\left|{\cal E}_{y}\left(0\right)\right|^{2} \geq\Delta_{0}$, as
\begin{equation}
s\left[p_{2}^{2}-\left(\frac{\omega}{c}\right)^{2}\Delta\right]^{1/2} + p_{2}=-\omega\mu_{0}\sigma^{\prime\prime},
\label{eq:dr2}
\end{equation}
where the real part of the graphene conductivity, $\sigma^{\prime}=\mathrm{Re}\left[\sigma\left(\omega\right)\right]\equiv 0$, has been neglected for clarity (this corresponds to the limit of a dispersive dielectric).
The above equation can be solved for $p_2$ and we obtain
\begin{equation}
p_2=-\frac{\Delta+c^{2}\mu_{0}^{2}\sigma^{\prime\prime2}}{2c^{2}\mu_{0}\omega^{-1}\sigma^{\prime\prime}}\:,
\label{eq:P2}
\end{equation}
which links $\omega $, $k_x$, and $\Delta $. One can consider $\omega $ and $\Delta $ as independently controlled parameters determined by the excitation conditions (in particular, $\Delta $ can be adjusted via the intensity of the incident wave).

Equation (\ref{eq:P2}) allows for a simple qualitative analysis. Requiring both the positiveness of $p_2$
and the condition that Eq. (\ref{eq:dr2}) also holds
one can
obtain the domains of existence of the TE-surface wave. The case $\sigma^{\prime\prime}>0$ is the simplest one. Here the
surface waves exist when
\begin{equation}
\Delta<-c^{2}\mu_{0}^{2}\sigma^{\prime\prime2}
\label{eq:Delta_neg}
\end{equation}
and $s=-1$.
We stress that this result is in contrast with the linear regime \cite{Mikhailov2007},
where only for $\sigma^{\prime\prime}<0$ TE-waves can propagate on a graphene-covered interface.
On the other hand,  for the case $\sigma^{\prime\prime}<0$ the surface waves exist in the domains,
\begin{equation}
-c^{2}\mu_{0}^{2}\sigma^{\prime\prime2}<\Delta<c^{2}\mu_{0}^{2}\sigma^{\prime\prime2}
\label{eq:s-positive}
\end{equation}
for  $s=1$ (this is the less interesting case),
and
\begin{equation}
\Delta>c^{2}\mu_{0}^{2}\sigma^{\prime\prime2}
\label{eq:Delta_pos}
\end{equation}
for $s=-1$. In conclusion, the existence domains for $s=-1$ are very wide.

\begin{figure}
\includegraphics[width=8.5cm]  {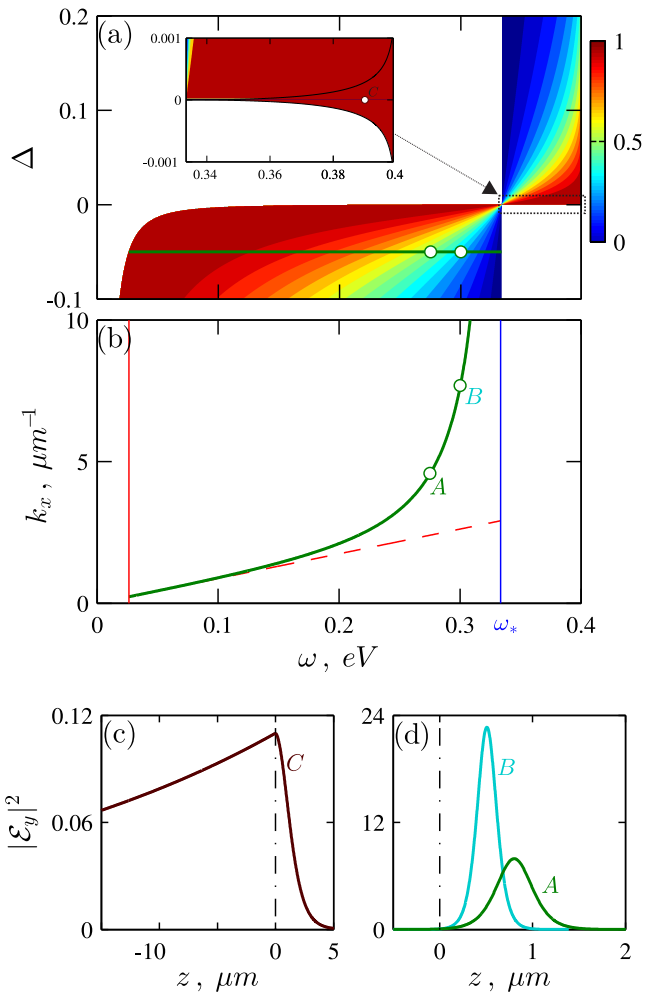} % {fig2_graphen}
\caption{ (Color online) (a) Surface polariton dimensionless phase velocity $\omega\sqrt{\varepsilon_{2}}/ck_x$ {\it versus} frequency and parameter $\Delta$. Inset in panel (a) depicts zoom in the vicinity of $\Delta=0$; (b) Dispersion relation $k_x\left(\omega\right)$ {[}solid line{]} for fixed $\Delta=-0.05$ [extracted from panel (a) along the respective horizontal line] and the light line in the linear dielectric $k_{x}=\left(\omega/c\right)\sqrt{\varepsilon_{2}}$  {[}dashed red line{]}; (c,d) Spatial profiles $\left|{\cal E}_{y}\left(z\right)\right|^{2}$ of surface polaritons for $\Delta=0$, $\omega=0.39\,$eV {[}curve C in panel (c){]} and  $\Delta=-0.05$, $\omega=0.275\,$eV, or $\omega=0.3\,$eV {[}curves A and B in panel (d), correspondingly{]}. The curves A--C correspond to the respective points in panels (a,b). The interface between the linear and nonlinear dielectrics is depicted by vertical dash-and-dotted black line. In all panels the dielectric permittivities of the nonlinear and linear media are $\varepsilon_{1}=2.89$, $\varepsilon_{2}=3$, while the relaxation rate and the chemical potential of graphene are $\Gamma=0$ and $\mu=0.2\,$eV.
\label{fig:e1m}
}
\end{figure}

\begin{figure} [t]
\includegraphics[width=8.5cm] {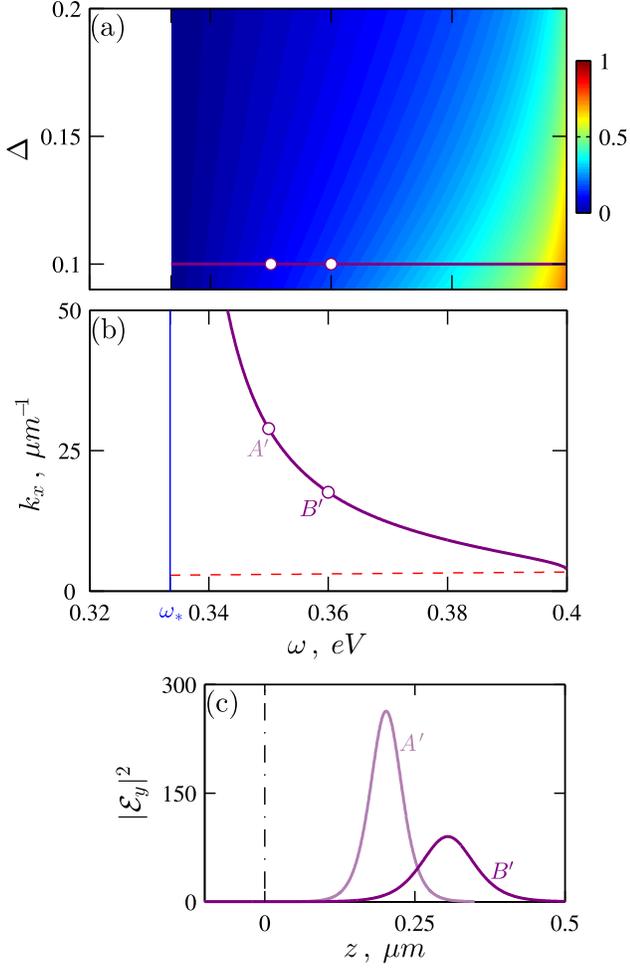} % {fig3_graphen}
\caption{ (Color online)
(a)  Surface polariton dimensionless phase velocity $\omega\sqrt{\varepsilon_{2}}/ck_x$ {\it versus} frequency and parameter $\Delta$; (b) Dispersion relation $k_x\left(\omega\right)$ {[}solid line{]} corresponding to the horizontal line $\Delta=0.1$ in panel (a) and the light line in the linear dielectric, $k_{x}=\left(\omega/c\right)\sqrt{\varepsilon_{2}}$  {[}dashed red line{]}; (c) Spatial profiles $\left|{\cal E}_{y}\left(z\right)\right|^{2}$ of surface polaritons for the parameters: $\Delta=0.1$, $\omega=0.35\,$eV, or $\omega=0.36\,$eV {[}curves A$^{\prime}$ and B$^{\prime}$, correspondingly{]}. The curves A$^{\prime}$ and B$^{\prime}$ correspond to the respective points in panels (a,b). The interface between the linear and nonlinear dielectrics is depicted by vertical dash-and-dotted black line. In all panels the dielectric permittivities of the nonlinear and linear media are $\varepsilon_{1}=2.89$, $\varepsilon_{2}=2.8$, while the relaxation rate and the chemical potential of graphene are $\Gamma=0$ and $\mu=0.2\,$eV.
\label{fig:e1g}
}
\end{figure}

The existence domains are depicted in Figs.~\ref{fig:e1m}(a) and Fig.~\ref{fig:e1g}(a) for the cases $\varepsilon_{1}<\varepsilon_{2}$ and $\varepsilon_{1}>\varepsilon_{2}$, respectively. In both
 plots, the white color represents domains where the
nonlinear surface polaritons cannot exist, while the bottom limits in these plots correspond
to the limiting value  $\Delta=\Delta_{0}$.

For $\varepsilon_{1}<\varepsilon_{2}$
{[}see Fig.~\ref{fig:e1m}(a){]} there are three existence domains:
(i) the finite domain at frequency range $\omega<\omega_{*}$, characterized
by positive $\sigma^{\prime\prime}>0$, negative $s=-1$ and restricted
by the curves (\ref{eq:Delta_neg}), $\Delta=\Delta_{0}$, and $\omega=\omega_{*}$;
(ii) the infinite domain at frequency range $\omega>\omega_{*}$, characterized
by negative $\sigma^{\prime\prime}<0$, positive $s=1$ and restricted
by the curve (\ref{eq:s-positive}) {[}see the inset in Fig.\ref{fig:e1m}(a){]};
(iii) the infinite domain at frequency range $\omega>\omega_{*}$, characterized
by negative $\sigma^{\prime\prime}<0$, negative $s=-1$ and restricted
by the curves (\ref{eq:Delta_pos}) and $\omega=\omega_{*}$. The last two regions are semi-infinite in frequency and $\Delta $.

For
$\varepsilon_{1}>\varepsilon_{2}$ {[}see Fig.\ref{fig:e1g}(a){]}
there is only one infinite  domain at frequency range $\omega>\omega_{*}$
(although the $s=1$ domain is also present here, for these particular
 parameters it is just a narrow strip in the vicinity of the frequency $\omega=2\mu$),
characterized by negative $\sigma^{\prime\prime}<0$, negative $s=-1$,
and restricted by the curves $\Delta=\Delta_{0}$, $\omega=\omega_{*}$,
and Eq. (\ref{eq:s-positive}).

The intensity scale in Fig.~\ref{fig:e1m}(a) and Fig.~\ref{fig:e1g}(a)
corresponds to the phase velocity $\omega/k_{x}$ (in units of the velocity
of light in the linear dielectric, $c/\sqrt{\varepsilon_{2}}$). The phase velocity attains its minimum value
in the vicinity of the frequency $\omega=\omega_{*}$ (where $\sigma^{\prime \prime}=0$) and decreases
with an increase of $\left|\Delta\right|$.
In  Figs.~\ref{fig:e1m}(b) and \ref{fig:e1g}(b) this behavior is seen in more detail, for the dispersion curves $k_{x}(\omega)$
obtained from Figs.~\ref{fig:e1m}(a) and \ref{fig:e1g}(a)
along the respective horizontal lines. From Figs.~\ref{fig:e1m}(b)
and \ref{fig:e1g}(b) it is seen that far from the frequency $\omega=\omega_{*}$
the wavevector $k_{x}$ is close to that in the linear dielectric, while
in the vicinity of $\omega=\omega_{*}$ the wavevector $k_{x}$
significantly exceeds its counterpart in the linear medium.

A very important piece of information is
the spatial profile of the electric field
$\left|{\cal E}_{y}\left(z\right)\right|^{2}$. It is  depicted in Figs.~\ref{fig:e1m}(c,d)
and \ref{fig:e1g}(c). Here, in the case $s=1$ {[}see Fig.~\ref{fig:e1m}(c){]}
surface polariton profile resembles its linear counterpart -- the
electric field maximum is at the interface between linear and nonlinear
media and the wave is poorly localized. Nevertheless, due to the nonlinearity of the substrate,
 the mode localization is much stronger for $z>0$ than in the linear dielectric. For the case  $s=-1$ {[}Figs.~\ref{fig:e1m}(d)
and \ref{fig:e1g}(c){]} the maximum of the spatial soliton is situated inside
the nonlinear dielectric, and the increase of $k_x$ shifts it closer to the inteface (smaller $\vert z_0 \vert$) and is
accompanied by an increase of the soliton amplitude and a decrease of its width (compare curves A and B, as well as curves A$^{\prime}$ and
B$^{\prime}$). These are two central results of this work: the
TE-type surface wave shows subwavelength localization and the intensity of the
field is quite high. Such a behavior is not found in the linear theory.

One of the advantages of graphene in electronics is the possibility
to tune its conductivity by electrostatic gating. So, a natural question is
how sensitive is the phase velocity of the nonlinear surface polariton with respect
to the variation of the graphene chemical potential?
We have found that for fixed values of the  parameter $\Delta$ and frequency $\omega$, an increase
of the chemical potential, $\mu$, results in an increase [when $\varepsilon_{1}<\varepsilon_{2}${]},
or decrease {[} when $\varepsilon_{1}<\varepsilon_{2}${]} of the phase velocity. Moreover,
varying the chemical potential, the phase velocity can be varied in a wide
range, from almost zero up to the velocity of light in the linear
dielectric, $c/\sqrt{\varepsilon_{2}}$.

\begin{figure} [t]
\includegraphics [width=8.5cm]  {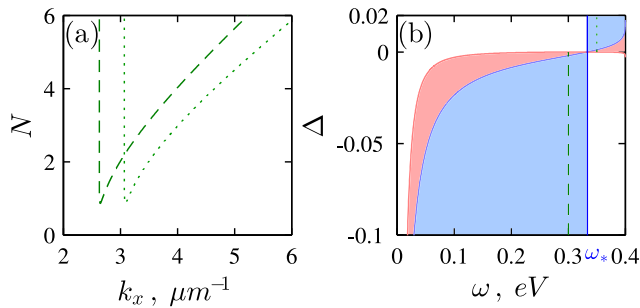}  % Fig4add_new.eps
\caption{ (Color online)
(a) Soliton norm $N$ {\it versus} wavevector $k_x$ for $\omega=0.3\,$eV (dashed line) and $\omega=0.35\,$eV (dotted line); % \textit{vs} for $\omega<\omega_*$ and $\omega>\omega_*$, respectively
(b) Regions of stability and instability of the nonlinear surface waves (domains shaded in blue and red, respectively) in the $(\Delta,\omega)$ plane. Dashed and dotted vertical lines correspond to the respective curves in panel (a).
The dielectric permittivities are $\varepsilon_1=2.89$ and $\varepsilon_2=3$.
\label{fig:sp-evol_prev}
}
\end{figure}

\textit{Stability of Nonlinear Waves.}
When dealing with the solutions of a given nonlinear problem, the central question concerns the stability of these solutions.
To address this issue, we introduce the norm, $N$, as a dimensionless integral:
\begin{equation}
%\begin(aligned)
N=\frac{\omega}{c}\int{\cal E}_{y}^{2}\left(z\right)dz=
\displaystyle{\frac{c}{\omega} \frac{(p_1+p_2)^2 - (\omega\mu_0\sigma^{\prime\prime})^2}{2p_2}}.
% \frac{c\left[p_1^2-(p_2+c\mu_0\sigma^{\prime\prime})^2\right]}{2\omega p_2}+\frac{c\left[p_1+p_2+c\mu_0\sigma^{\prime\prime}\right]}{\omega}
\label{eq:norm}
%\end(aligned)
\end{equation}
In the case $\varepsilon_1<\varepsilon_2$ (i.e. $\Delta _0<0$, the norm for a fixed $\omega$ has a minimum
[see Fig.~\ref{fig:sp-evol_prev}(a)] at the critical wavevector,
\begin{eqnarray}
& &k_{x_{\text {crit}}}^2=\frac{\omega^2}{c^2}\left \{\varepsilon_2 - \frac{\Delta _0+(c \mu_{0} \sigma^{\prime\prime})^2}{8(c \mu_{0} \sigma^{\prime\prime})^2}\times \left\{-3\Delta _0+(c \mu_{0} \sigma^{\prime\prime})^2\right. \right. \nonumber\\
& &\left. \left. - \sqrt {\left[\Delta _0+(c \mu_{0} \sigma^{\prime\prime})^2\right]\times \left[9\Delta _0+(c \mu_{0} \sigma^{\prime\prime})^2\right]}\right\}\right\}.\label{eq:Kcrit}
\end{eqnarray}
Moreover, if $(c \mu_{0} \sigma^{\prime\prime})^2 \ll \varepsilon_2-\varepsilon_1$, the expression for the
critical wavevector (\ref{eq:Kcrit}) reduces to
\begin{equation}
k_{x_{\text {crit}}}^2\approx  \left(\frac{\omega}{c} \right)^2 \displaystyle{\left[\varepsilon_2 - \frac{\Delta _0+(c \mu_{0} \sigma^{\prime\prime})^2}{3} \right]}.\label{eq:Kcrit1}
\end{equation}
In the stability analysis, the critical wavevector $k_{x_{\text {crit}}}$ plays an important role. It determines the boundary in the domain of parameters $\Delta$ and $\omega$ corresponding to stable and unstable modes. In other words, nonlinear polaritons with $k_x<k_{x_{\text {crit}}}$ (for which $\partial N/\partial k_x<0$) are unstable~\cite{Akhmediev}, while in the opposite case of $k_x>k_{x_{\text {crit}}}$ ($\partial N/\partial k_x>0$) the nonlinear surface wave is stable (according to the conventional Vakhitov-Kolokolov criterion for the soliton stability~\cite{c:Vakh}). We point out that the presence of graphene (with $\sigma^{\prime\prime}\ne 0$) results in lowering the critical wavevector value (\ref{eq:Kcrit1}), in comparison with the case without graphene (considered in Ref.~\cite{Akhmediev}).

\begin{figure} [hbt]
\includegraphics [width=8.5cm] {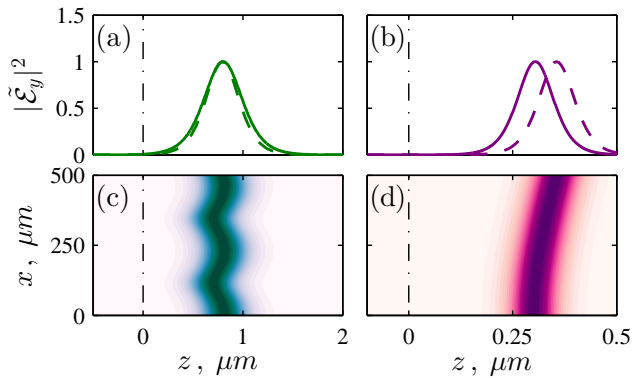} % {fig4_graphen}
\caption{ (Color online) Evolution of the spatial profile of the nonlinear surface polariton as it propagates along $x$. In panels (a) and (b), solid and dashed lines correspond to the initial (at $x=0\,\mu$m) and final (at $x=500\,\mu$m) wave profiles, respectively.
Parameters of panels (a,c) and (b,d) correspond to points A in Fig.~\ref{fig:e1m} and B$^{\prime}$ in Fig.~\ref{fig:e1g}, respectively. Shown is the beam intensity  $\left|\tilde{\cal E}_{y}\left(x,z\right)\right|^{2}$ normalized to unity. In all panels the dash-dotted line marks the position of an interface between the linear and nonlinear dielectrics where the graphene layer is located.
\label{fig:sp-evol}
}
\end{figure}

The stable and unstable domains are depicted in Fig.~\ref{fig:sp-evol_prev}(b). We see that for $\omega<\omega_*$ (when $\sigma^{\prime\prime}>0$) the nonlinear modes are unstable in the vicinity of the domain boundary (\ref{eq:Delta_neg}), whereas for $\omega>\omega_*$ (when $\sigma^{\prime\prime}<0$) unstable region includes the \textit{full} domain (\ref{eq:s-positive}) as well as part of domain (\ref{eq:Delta_pos}) in the vicinity of its boundary. In other words, all the nonlinear surface polaritons with $s=1$ are unstable. Fortunately, this case is not interesting since
it corresponds to weekly confined nonlinear TE-waves. However, both polaritons depicted in Fig.~\ref{fig:e1m}(d)
{[}$\varepsilon_{1}<\varepsilon_{2}$ and $\sigma^{\prime\prime}>0${]} are stable
and this has been confirmed by numerical integration of Eq.~(\ref{eq:wave-eq}) [see Figs.~\ref{fig:sp-evol}(a) and \ref{fig:sp-evol}(c)].

In the opposite case {[}$\varepsilon_{1}>\varepsilon_{2}$ and $\sigma^{\prime\prime}<0${]} the norm $N$ of the localized waves does not have a minimum and grows monotonically with the increase of $k_x$, i.e. we have the stability of the nonlinear polaritons in the full existence domain. Numerical integration of Eq.~(\ref{eq:wave-eq}) shows [see Figs.~\ref{fig:sp-evol}(b) and \ref{fig:sp-evol}(d)] that the nonlinear wave is not collapsing and maintains its shape. In this sense, the Vakhitov-Kolokolov criterion is not violated here. At the same time, the nonlinear wave undergoes a drift instability~\cite{Sivan}, so that in the course of propagation the center-of-mass of the spatial soliton in the nonlinear medium gradually moves away from the interface $z=0$.

\textit{Conclusions.}
We have studied analytically and numerically the propagation of electromagnetic waves
along the surface of a nonlinear dielectric medium covered by a layer of graphene.
We have shown that the presence of a single graphene layer leads to the existence and stabilization
of nonlinear surface modes with the maximum amplitude located either at the interface or inside of
the nonlinear dielectric medium. We also found the stability domains of these nonlinear modes.

\textit{Acknowledgements.}
Y.V.B. thanks Nonlinear Physics Center at the Australian National University for a warm
hospitality during his visit at the initial stage of this work. This work was partially
supported by the FEDER COMPTETE Program and by the Portuguese Foundation for Science
and Technology (FCT) through grant PEst-C/FIS/UI0607/2013. The authors thank I. Iorsh
and I. Shadrivov for useful discussions.

% \bibliographystyle{apsrev}
% \bibliography{sp_graphene_s}

\end{document}